# Neural representation in active inference: using generative models to interact with—and understand—the lived world

Giovanni Pezzulo[1,*], Leo D'Amato[1,2], Francesco Mannella[1], Matteo Priorelli[3], Toon Van de Maele[4], Ivilin Peev Stoianov[3], Karl Friston[5,6]

1. Institute of Cognitive Sciences and Technologies, National Research Council, Rome, Italy
2. Polytechnic University of Turin, Turin, Italy
3. Institute of Cognitive Sciences and Technologies, National Research Council, Padua, Italy
4. IDLab, Department of Information Technology, Ghent University - imec, Ghent, Belgium
5. Wellcome Centre for Human Neuroimaging, Queen Square Institute of Neurology, University College London, London, UK
6. VERSES Research Lab, Los Angeles, CA, USA

**Corresponding author:**
Giovanni Pezzulo,
Institute of Cognitive Sciences and Technologies,
National Research Council,
Via S. Martino della Battaglia 44,
00185, Rome, Italy.
Giovanni.pezzulo@istc.cnr.it

**Abstract**

This paper considers neural representation through the lens of active inference, a normative framework for understanding brain function. It delves into how living organisms employ generative models to minimize the discrepancy between predictions and observations (as scored with variational free energy). The ensuing analysis suggests that the brain learns generative models to navigate the world adaptively, not (or not solely) to understand it. Different living organisms may possess an array of generative models, spanning from those that support action-perception cycles to those that underwrite planning and imagination; namely, from "explicit" models that entail variables for predicting concurrent sensations, like objects, faces, or people—to "action-oriented models" that predict action outcomes. It then elucidates how generative models and belief dynamics might link to neural representation and the implications of different types of generative models for understanding an agent's cognitive capabilities in relation to its ecological niche. The paper concludes with open questions regarding the evolution of generative models and the development of advanced cognitive abilities – and the gradual transition from "pragmatic" to "detached" neural representations. The analysis on offer foregrounds the diverse roles that generative models play in cognitive processes and the evolution of neural representation.

**Introduction**

*"My thinking is first and last and always for the sake of my doing."*
—William James

The concept of "neural representation," which pertains to the idea that the brain represents elements of the external world, plays a prominent role in neuroscience, psychology, and philosophy. However, its interpretation remains a subject of discussion [1–9]. The notion of representation is intricate, particularly when applied to brain science, where it necessitates a connection to specific brain activities or states. This engenders questions such as: What kind of neural state or activity would count as a "representation", and how? What would be the entity out there that is "represented"? Does the brain really represent something or is the notion of neural representation misplaced?

In this article, we approach the concept of neural representation through the lens of active inference: a normative framework for describing brain function and cognitive processes [10]. Active inference proposes that the brain constructs a generative model encompassing the external world, the body, and action possibilities. This model underwrites sense making and purposeful interactions with the surroundings. For instance, it enables the apperception of the visual image in Figure 1, while also offering various affordances (e.g., catching the fishes — or not, in this instance).

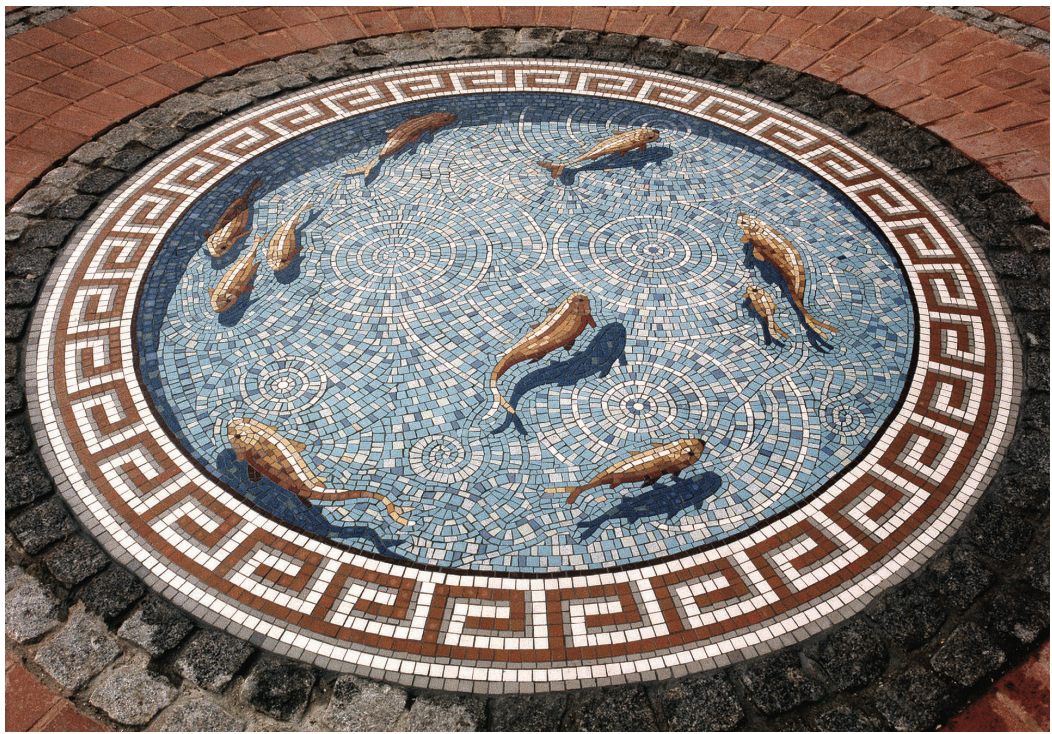

***Figure 1.*** *Fishpond Mosaic designed by Gary Drostle and made by Gary Drostle and Rob Turner in 1996 for Croydon Council, south London. Source: Wikipedia.*

The connection between active inference and diverse philosophical concepts of representation has triggered substantial discussion. For instance, it intersects with internalist perspectives [11] and action-centered viewpoints [12], alongside enactive theories that negate the idea of representation [13]. Adding complexity to the debate is the variance in implicit notions of representation among researchers, each accentuating distinct criteria [14].

Here, we sidestep this debate and focus instead on how neural representations might emerge from the reciprocal exchanges between living organisms and the niches they inhabit; particularly, when viewed through the lens of active inference. Our focus is on elucidating the formal constructs employed by active inference to explain how living organisms solve cognitive problems. These constructs encompass generative models, probabilistic beliefs, and the concept of variational free energy. By delving into these constructs, we hope to naturalize the notion of neural representation under a first principles account of sentient behavior.

**Surprise minimization, active inference and action-perception loops**

Each living entity confronts the inherent challenge of adaptively regulating and managing essential parameters crucial for its existence, such as body temperature or glucose levels. From a formal perspective, the imperative for adaptive control can be conceptualized remaining in a limited array of potential states, or to inhabit an organism-specific niche that satisfies essential needs. An example of this imperative is the requirement to maintain body temperature at approximately 37°C. In terms of information theory, states that deviate from these acceptable bounds are deemed "surprising". For instance, significantly higher or lower sensed body temperatures than 37°C fall under this technical concept of surprise (i.e., surprisal or self-information). This underscores the vital importance for a living organism to minimize the surprise inherent in its sensory interactions with the environment.

The notion of surprise, in turn, depends on two key factors: actively sampled sensory *observations* from the external world and internally generated *predictions* regarding these observations. More precisely, surprise increases with the discrepancy between predictions and observations, see Figure 2A. An organism can mitigate this discrepancy in two complementary ways: by adapting its predictions to anticipate forthcoming observations (perception) more accurately or by influencing its surroundings to ensure alignment of forthcoming observations with predictions (action).

Active inference formalizes the reduction of surprise as the minimization of variational free energy—a quantity that places an upper bound on surprise [10]. The two constituents of this measure—divergence and evidence—correspond to perception and action, respectively, see Figure 2B. Under certain simplifications, variational free energy reduces to (precision weighted) prediction error. This will prove significant in our subsequent discussion of predictive coding.

The interaction between action and perception is vital to reduce the gap between what an animal predicts and what it actually experiences. This difference is measured as variational free energy. Having accurate perception is important because if an animal can't sense its surroundings correctly, it might struggle to choose the right actions. Likewise, if it doesn't take actions to satisfy its basic needs like thirst, it might not survive for long.

Importantly, action is guided by an inherent optimism, described as a pre-existing preference (or just "prior") for outcomes that ensure survival; namely, the outcomes or states of being that are characteristic of the agent in question. Consider a creature under the midday sun. In such a scenario, if its predictions encompassed outcomes like sunburn and dehydration, it would fare poorly. Instead, a more successful animal might predict a constant temperature of 37°C, prompting it to seek out shade [15]. This illustration underscores that within active inference, certain priors, such as "my body temperature is around 37°C", extend beyond the conventional scope in Bayesian statistics—typically indicating a priori knowledge about the external environment—and instead, assume a role akin to 'set points' in cybernetics. These priors encode task objectives as attractors within the organism's state space, enabling error correction and negative feedback control [16]. That's because any difference between the expected and currently sensed body temperature would trigger adaptive adjustments (like vasodilation or finding shelter) to resolve the difference.

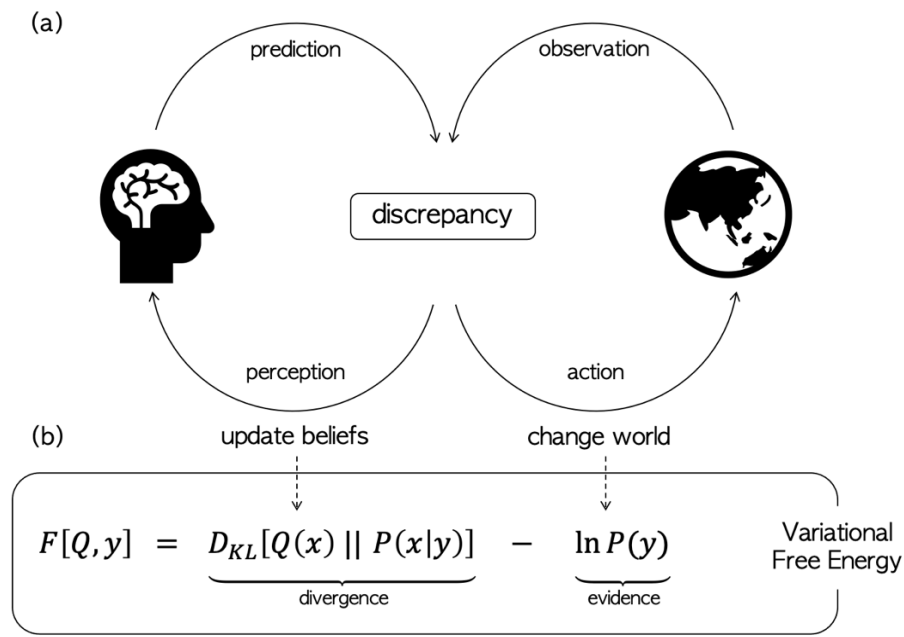

*Figure 2. Complementary Roles of Action and Perception in Variational Free Energy Minimization.* *(a) Action and perception play distinct yet interdependent roles in minimizing the discrepancy between an organism's predictions and the world. Perception reduces this discrepancy by shaping the organism's beliefs to align more closely with the world, enhancing predictive accuracy. On the other hand, action diminishes the discrepancy by modifying the external environment, making it better match the organism's beliefs. (b) Variational free energy mathematically captures the concept of "discrepancy". While this paper does not delve into the technical details of variational free energy minimization, it encompasses two components— divergence and evidence—both addressed by perception and action, respectively. Further discussion can be found in [10].*

The same feedback loop applies to controlling movement. Figure 3 gives a simple instance of active inference in posture control for a "finger." The finger's posture is determined by a single factor: its angle. Figure 3A displays perceptual inference—inferring the finger's angle from what is sensed. In this basic setup, the influence of both the prior (preferred angle) and action is absent, so the finger remains stationary. Initially, the actual angle of the finger (dark blue) is vertical, while the inferred angle, the average angle in the agent's beliefs (light blue), is mistakenly set as horizontal. The system's internal feedback doesn't match the inferred angle, leading to a difference (proprioceptive prediction error: **propr pred err**, shown in green) between the real and inferred angles. The system minimizes variational free energy (in this case, minimizing prediction error) by "changing its mind", and after around 100 steps, the inferred angle aligns with the actual angle.

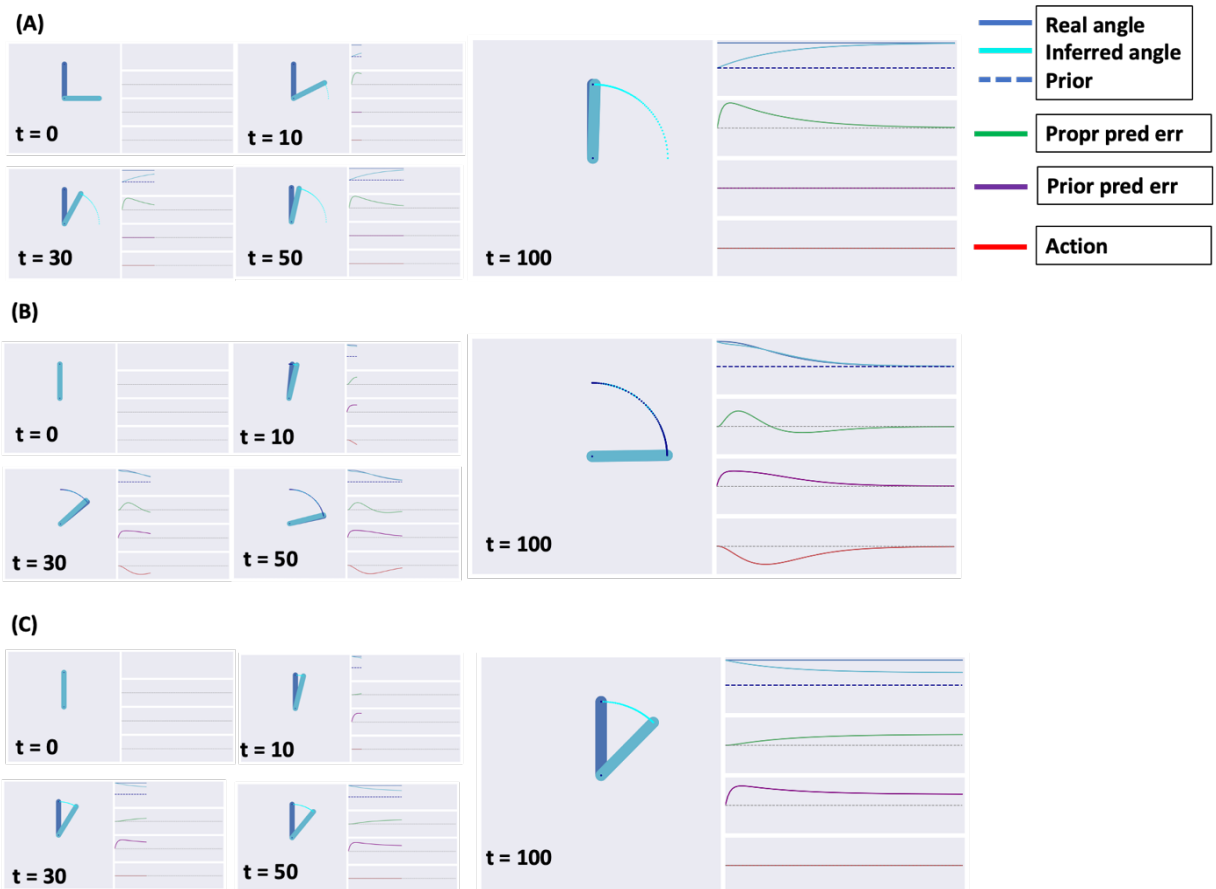

*Figure 3. Illustration of Active Inference for Postural Control of a "Finger". (A) Perception serves to minimize the discrepancy between the actual angle of the finger in the generative process (represented by the dark blue line) and the estimated angle, which corresponds to the agent's posterior belief (depicted in light blue). (B) Collaboratively, perception and action work to diminish the gap between the preferred angle (prior belief indicated by the dotted blue line) and the presently inferred angle (depicted in light blue) of the finger by adjusting the actual finger angle (dark blue). This alignment results from the finger being moved toward the preferred angle. (C) In the absence of action but when the prior belief is factored in, the inferred angle converges to an intermediate point between the actual position and the prior belief. The "t" symbols signify discrete time steps in the simulation. Refer to the main text for a comprehensive explanation.*

Figure 3B serves as a follow-up to Figure 3A, introducing the influences of both the prior and action. Here, the image exemplifies the cooperative dynamic between action and perception, guiding movement toward the preferred (horizontal) angle. Initially, the finger's real and inferred angles align with the vertical position. However, the system leans towards a prior preference for a horizontal finger angle, predicting it consistently. As a result, there's a mismatch (*prior pred err*, shown in violet) between the prior and inferred angles. This prompts the inferred angle (light blue) to shift towards the favored (prior) angle, leading to a second discrepancy (*propr pred err*, depicted in green) between the real and inferred angles. This time, the difference can be bridged by acting (red line), by physically moving the real finger (dark blue). The system reduces variational free energy by merging action and perception, causing both real and inferred angles to converge with the preferred (prior) angle, after about 100 steps. Notably, active inference differentiates from predictive coding in that it allows for acting on the world to fulfill predictions, such as moving to a preferred state. It's worth mentioning that this is a simple example. Later we will see how this framework can be extended to encompass a more complex notion of purposeful action—where organisms can envision future movement targets, establish them as priors, and finally attain them through planned action.

Lastly, Figure 3C offers an alternative extension of Figure 3A, incorporating the influence of the prior while excluding action. Here, following Bayes' rule, observations and priors are merged, causing the inference of the angle's position to gravitate toward a midpoint between the real position and the prior belief. (note that in this simulation, both prior and observations carry the same weight. In active inference, this weight is contingent upon the precision or reciprocal variance of the comparative information streams—a topic we will address in more depth later.)

To recap, we have introduced the fundamental imperative for living organisms as the minimization of the discrepancy between predictions and observations (technically, variational free energy). Additionally, we clarified that both perception and action play a concerted role in achieving this minimization. This viewpoint underscores the brain's nature as a predictive apparatus, continually generating predictions about the external world to steer both perceptual understanding and action regulation. This stands in contrast to viewing the brain as a mechanism primarily focused on converting external stimuli into internal representations and subsequent motor responses. However, we have yet to elaborate on the neural mechanisms that living entities employ to formulate their predictions and conduct their inferences. This brings us to our next section.

**Generative models**

How do organisms generate their predictions? In order to generate predictions and draw inferences, organisms rely on acquired generative models of the causes of their sensations. A generative model is a statistical description of how observations stem from unobserved (hidden or latent) states. For instance, it explains how a visual object, like an apple, produces an image on the retina. This concept of a generative model aligns with the influential perspective in cognitive science, suggesting that living entities carry small-scale models or cognitive maps of the external world [17,18]. It further resonates with the notion of "world models" in artificial intelligence [19,20].

While generative models are often associated with "models of the external world," this notion extends to encompass a broader scope: it includes models not only of the external world but also of the body (e.g., the "body schema"), the internal milieu (e.g., the "interoceptive schema"), emotional aspects, social dynamics, self-related constructs, and more [21–27]. The crucial point is that for survival, the generative models for living entities must not only provide ways to comprehend the world but also prescribe methods for action – thus attributing agency to organisms. In other words, the brain learns generative models to interact with the world, not (or not just) to understand it. Consequently, models that include potential actions and implicit affordances [28] become pivotal in the context of living organisms.

Figure 4 illustrates the key distinction between the brain's "generative model" of the external world and the actual external world (referred to as the "generative process" responsible for an organism's observations). The figure highlights several key aspects. Firstly, a generative model captures the statistical relations between observables (*y*) and states (*x*) that are hidden from direct observation. This equates formally to the joint probability distribution over unobservable causes and observable consequences *P(x,y)*. Secondly, it's crucial to note that a generative model doesn't equate to reality itself; rather, it represents an organism's interpretation of reality — which it can realize or author through action. Both the generative model and the generative process incorporate "hidden states," but these states need not necessarily correspond or even resemble each other. The hidden states within the organism's generative model (*x*) support Bayesian beliefs, representing probability distributions over latent states that are used to predict sensory consequences, including, crucially, the consequences of our movements and physiology that can be realized by motor and autonomic reflexes, respectively.

However, the hidden states used to generate predictions are not isomorphic with hidden variables in the external world (*x\**) — they can be distinct variables, such as categorical versus continuous. For example, I may have the Bayesian belief that I am writing this sentence. However, in reality, my movements are the result of electromechanical forces exerted by my musculature that are trying to minimize proprioceptive prediction errors; thereby fulfilling the proprioceptive predictions entailed by my beliefs (much in the spirit of ideomotor theory). This distinction between *internal* hidden states (of the generative model) and *external* hidden states (of the generative process) gains significance as we deconstruct the concept of neural representation.

The relationship between the agent and the world is interactive, forming loops between action and perception. The generative model influences the generative process via actions, while the generative process shapes the generative model through observations. This dynamic implies that internal hidden states and external hidden states are statistically separate and cannot directly impact each other. This statistical separation is formally articulated through the concept of a Markov blanket. Those interested in exploring this concept and its potential ramifications for the notion of representation may refer to [11].

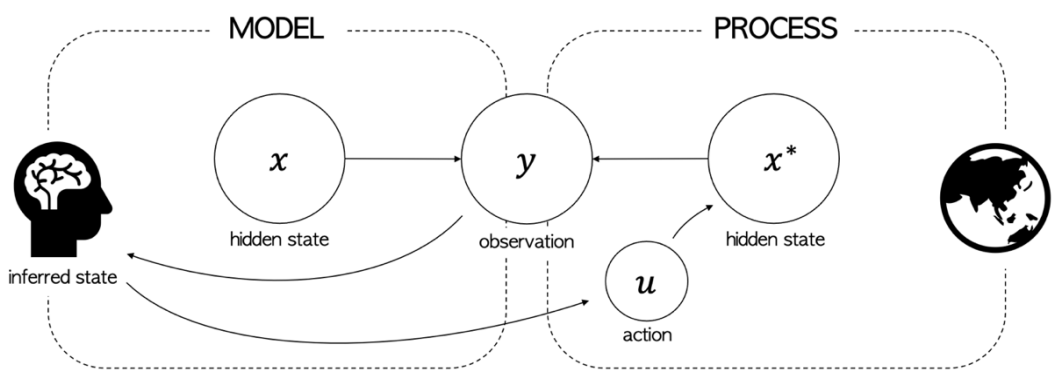

*Figure 4. Differences between the Generative Model and the Generative Process in Active Inference. Under active inference, the generative model and the generative process encompass distinct sets of hidden states (x and x\* respectively). These two components form interconnected action-perception loops: the generative model can exert influence over the generative process through action (u), while the generative process can affect the model through observations (y). The generative model is expressed mathematically as the joint probability distribution P(x,y). This model can generate hypothetical observations — i.e.,*

*predictions — (y) based on an inferred state (x). The generative model can be "inverted" for Bayesian inference – estimating the value of x after observing y (which may itself have been caused by the estimated value of x, as a self-fulfilling prediction). The outcome of this inference is a posterior or Bayesian belief. To update Bayesian beliefs, active inference employs a variational approximation, leading to an approximate posterior belief denoted by Q(x), which approximates the true posterior, labeled as P(x|y). See main text for further elaboration.*

An essential point to note is that generative models can be employed in two distinct directions: from x to y and *vice versa*. In the first direction, known as the *generative* direction, these models facilitate the generation of potential observations from inferred hidden states. Essentially, this means that the models can be used to create predictions and imaginative content, resulting in the designation "generative" models. This capacity is notably demonstrated by contemporary generative AI models that capture statistical patterns from extensive curated datasets and subsequently produce synthetic text (e.g., BERT, GPT) or images (e.g., Midjourney, DALL-E) [29–31]. When employed in the second direction, called the *inferential* direction, from y to x, generative models enable the inference and optimization of probabilistic beliefs about the hidden state of the world based on observations. This process occurs for example during perceptual inference, where the model infers the hidden state from observed data. Inference or Bayesian belief updating corresponds to inverting the generative model in the inferential direction.

The dual inferential and generative roles of generative models furnish mechanistic insights into the fundamental cognitive functions of the brain. The processes of perception (via predictive coding) and action (via active inference) can be effectively described in terms of inference. On the other hand, tasks such as planning, and imagination necessitate the generation of anticipatory (hypothetical or counterfactual) observations. In the following section, we delve into illustrative instances of generative models that underpin these cognitive capacities, exploring their implications for the concept of neural representation.

**Generative models for action-perception loops**

Both perception and action can be effectively conceptualized as inference processes. The concept of perception as unconscious inference traces back to [32]. Figure 5 illustrates a biologically inspired generative model that supports perceptual learning and inference: a hierarchical predictive coding network [33,34]. The hierarchical structure is discernible in the figure's layout, where neural populations are grouped from bottom (lower hierarchical level) to top (higher hierarchical level). Notably, each hierarchical level encompasses distinct neural populations that encode expectations (in deep cortical layers) and prediction errors (in superficial cortical layers). The "expectation" nodes encode probabilistic beliefs or their sufficient statistics (such as mean and variance for Gaussian distributions), which generate predictions regarding observed content. Conversely, the "prediction error" nodes capture the difference between predictions and observations. The process of inferring the most plausible explanation for observations unfolds through the minimization of prediction errors, enabled by reciprocal message exchange between units: top-down messages convey predictions, while bottom-up messages convey prediction errors. Learning operates similarly, driven by prediction error minimization: please see [33,34] for details.

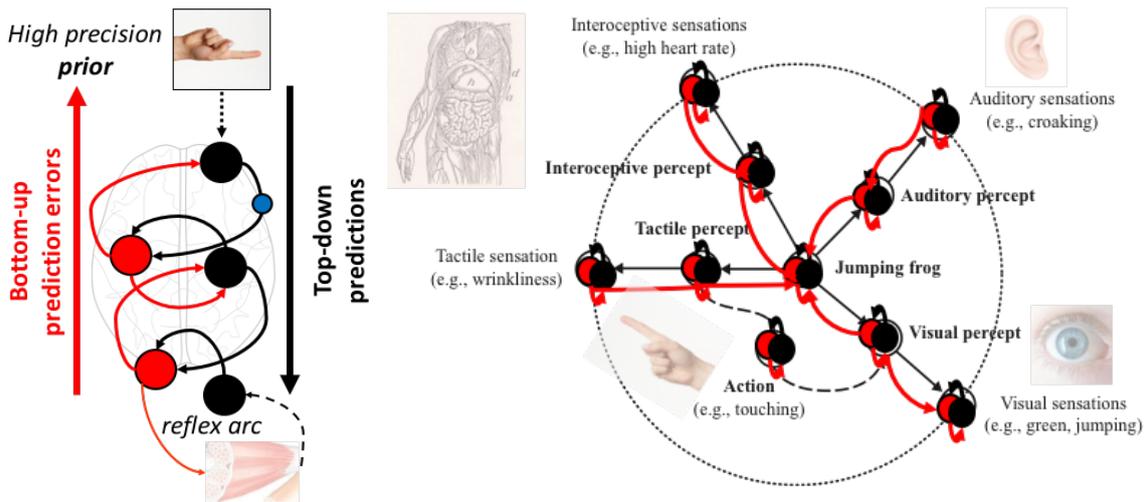

*Figure 5. Predictive Coding and Active Inference Models for Action-Perception. (A) Illustration of active inference as "predictive coding with motor reflexes" for movement control. Here, black and red nodes denote expectation and prediction error nodes, respectively. Top-down and bottom-up edges convey predictions and prediction errors, respectively, across levels, forming a predictive coding hierarchy. Prediction error can be minimized by either revising prior beliefs or (if these priors are held with high precision, as indicated in the figure by the blue node), by acting in the world. See the main text for explanation. (B) Hierarchical generative model designed for the (multimodal) notion of a jumping frog. This architecture frames the concept of a jumping frog as the underlying cause of numerous sensory outcomes, spanning visual cues of something leaping and green, and auditory inputs of croaking sounds. Some of these consequences can be contingent on actions, such as the visual input changing when the frog is foveated. The inversion of this generative model supports perceptual inference (e.g., recognizing a jumping frog) from its observed sensory manifestations (e.g., seeing something green and leaping), incorporating information from multiple sensory modalities.*

The example presented in Figure 5A revisits the process of inferring the "finger angle," as previously explained in Figure 3, through the lens of a predictive coding model. In this instance, the (posterior) belief regarding the finger angle is encoded in the expectation node at the lower hierarchical level, while its corresponding prior belief resides in the expectation node at the level above. These two nodes are connected via prediction error nodes that convey discrepancies with proprioceptive sensations and prior predictions. The lower section of Figure 5A illustrates the simplest manifestation of active inference, which demonstrates the potential to engage in action (to counteract proprioceptive prediction errors) by merely equipping a predictive coding network with a peripheral motor reflex arc [35,36]. This predictive coding architecture can be readily expanded by incorporating additional hierarchical levels. For instance, predictive coding models for visual processing, such as those involving natural images or MNIST letters, incorporate multiple hierarchical levels, where the expectations at each level encode increasingly abstract features of the stimulus [33,34,37].

Figure 5B offers a more sophisticated demonstration of a generative model, featuring a multimodal concept—a jumping frog. The image presents a hierarchical generative model, where peripheral nodes situated along the dotted circle denote sensations spanning diverse modalities (such as visual, auditory, and interoceptive). Nodes within the circle are unimodal hidden states, while the central node is a multimodal hidden state. The notion of a jumping frog encompasses a set of multimodal predictions: predictions about how a frog would appear, the sounds it would emit, and the typical narratives involving frogs, for instance. Action is part and parcel of this concept, enabling predictions about interactions with a frog (e.g., the sensations induced by palpating a frog; either by touching it or visually through foveating it): in brief, all the affordances offered by a frog. These predictions are grounded in the multifaceted experience of interacting with frogs, in various incarnations. The concept of a jumping frog encompasses perceptual inference—recognizing a frog or comprehending a dialogue about one—active inference—planning how to palpate a frog—and the ability to "mentally navigate" through the concept of jumping frogs, generating hypothetical, "as-if" predictions regarding frogs, for instance.

*Generative models, belief dynamics, and neural representation*

With the introduction of a basic generative model for active inference and its inferential dynamics (based prediction error minimization), we now explore the connection to neural representation. Three key points warrant attention in this regard. Firstly, there exists a consistent functional relationship between the (internal) hidden states within the hierarchical generative model and the (external) hidden states in the world, like the "finger" angle or MNIST letters. This connection is generally assumed (though not universally) to reflect the most commonly embraced notion of neural representation in cognitive psychology and neuroscience. Key aspects of this concept revolve around the structured nature of the connection, indicating a causal link between internal and external hidden states, as opposed to a simple correlation. This underscores that internal hidden states encode beliefs *about* external hidden states, even though a direct one-to-one mapping may be lacking. Furthermore, these hidden states *serve as tools for* the organism to steer adaptive prediction and control, holding meaning, significance and adaptive value for it in the process [5,7].

Secondly, it is crucial to distinguish between two concepts—generative models and probabilistic beliefs—often linked to the idea of neural representation, though they manifest through distinct neural mechanisms. The structure and parameters of the generative model, or its hierarchical components, could be encoded in synaptic connectivity. This is where elements like statistical regularities (like priors and likelihood functions) find their coding. On the other hand, probabilistic beliefs (or their statistics, such as the mean and variance of Gaussian distributions) might find their encoding in neuronal activity and the collective dynamics of neuronal populations. Over time, neuronal activity reflects Bayesian belief updating under the generative model. Significantly, the organism's (posterior) beliefs stem not solely from immediate sensations, but from an inference process operating under the generative model. These beliefs – which can be associated with neural representations – essentially blend sensory input, memory, and predictions, rather than merely *representing* sensory input. As highlighted in the example of the frog, generative models additionally enable what-if predictions about hypothetical outcomes of actions. We will revisit this point later when we discuss planning (as inference) and imagination.

Finally, it is important to recognize that the manner in which (predictive) coding occurs and the dynamics of belief updating are profoundly influenced by the specific generative model in play. Here, we have focused on simple generative models in the context of predictive coding, which unfold in continuous time and encode variables—such as the finger angle in Figure 4—as Gaussian distributions or their sufficient statistics (mean and variance). Nevertheless, a broader realm of generative models could be found within the brains of both basic and more advanced living creatures. These generative models might operate in continuous or discrete time, or both. They could be structured into distinct factors that autonomously impact observations (think of "where" and "how" in visual pathways). These models might also feature hierarchical organization with temporal depth, involving hidden states about both the past and future. Despite these variations, the overarching belief dynamics adhere to a common principle: the minimization of free energy or (precision weighted) prediction errors. Different generative models necessitate the transmission of "neural messages" (like predictions and prediction errors) among nodes in distinct manners. Consequently, they result in different patterns of connectivity and belief dynamics [10]. For instance, generative models explicitly encompassing past and future hidden states enable the exchange of messages "from the past" and "from the future" to update beliefs concerning "the present." In the following sections, we provide a brief overview of some illustrative examples and consider their standing in relation to neural representation.

**Generative models to solve cognitive tasks that involve planning and imagination**

Hitherto, our discourse has revolved around generative models that support perceptual and active inference through the minimization of variational free energy. In this section, we consider more expressive generative models, which have been employed to describe cognitive tasks involving planning and imagination. These tasks demand generative capacities and the ability to envision forthcoming scenarios and future observations.

*Hierarchical generative model for memory-guided spatial alternation*

Consider a two-level hierarchical generative model designed to solve memory-guided spatial alternation tasks [38]. These tasks require coordinated interactions within hippocampal (HC) and prefrontal circuits (mPFC). A depiction of a memory-guided spatial alternation task is presented in Figure 6A. In this scenario, within a W-maze arrangement, an organism (in this case, a simulated rodent) must navigate alternating corridors adhering to a learned rule (like center, left, center, right, center) to attain rewards [39,40].

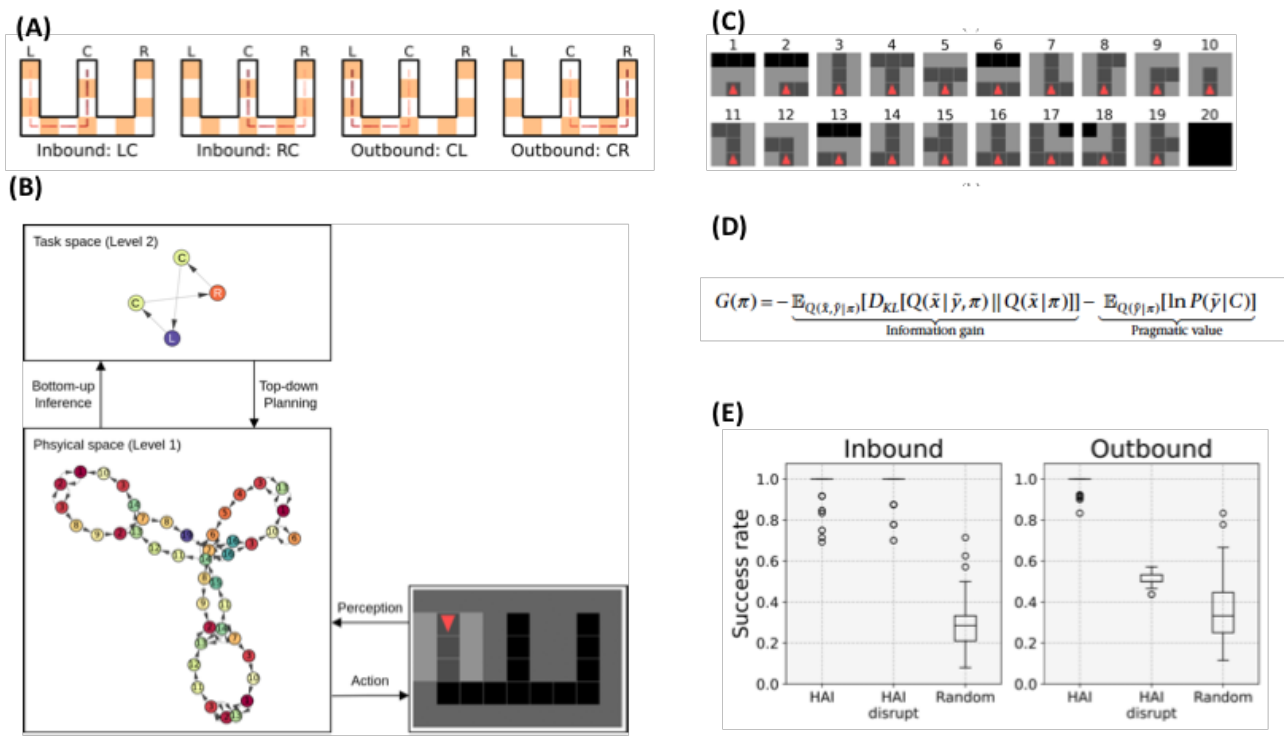

*Figure 6. Hierarchical Generative Model for Memory-Guided Spatial Alternation Tasks, adapted from* [38]. *(A) Depiction of the spatial alternation task within the W-maze used by* [39,40].

*(B) Hierarchical generative model used to solve the spatial alternation task. This model features two interconnected levels, each learning maps of physical space (level 1) and task space (level 2), supporting hierarchical perceptual inference, planning, and action-perception loops. (C) Collection of sensory observations encountered by the simulated rodent during navigation. (D) Planning involves the selection of policies (action sequences, π) by evaluating their expected free energy, denoted as G(π). Notably, the expected free energy considers forthcoming observations (anticipated by pursuing particular policies) beyond current and past observations, in contrast to variational free energy in [10] (E) Simulation outcomes illustrating the model's accuracy in spatial alternation performance. The results show that an intact (HAI) model correctly solves the task. However, hindering communication between the two levels of the hierarchical generative model impacts decision-making in challenging (outbound) decisions (HAI-disrupt). Please see [38] for details.*

The hierarchical generative model shown in Figure 6B effectively solves the spatial alternation task. In the process of exploring the maze, the two levels of this model learn "cognitive maps" to handle both physical space (HC map [41]) and task space (mPFC map [42]). Here, the term "cognitive map" signifies a set of hidden variables united into a coherent structure, often resembling a graph. This structure encodes the relationships between spatial positions within a maze (or stages of a task) and facilitates a form of "mental navigation." For instance, it allows predictions about the potential outcomes of traversing the maze (or progressing to the next task stage).

The cognitive map of physical space encompasses the maze's layout and the feasible actions within it. This map is constructed by learning the statistical relationships between sequences of sensory observations encountered during navigation (as depicted in Figure 6C). Notably, some observations are "aliased," meaning they appear identical in multiple maze segments, contributing to self-localization ambiguity. To address this, a statistical sequence learning algorithm known as a clone structured cognitive graph (CSCG) [43] is employed. This algorithm resolves ambiguity and encodes trajectories as sequences of hidden states that mirror the animal's positions. The outcome is a cognitive map that closely resembles the physical space (as depicted in the Level 1 box of Figure 6B).

The cognitive map of task space captures the spatial alternation rule (e.g., center, left, center, right, center). This map is learned on top of the cognitive map of physical space. To create it, statistical dependencies between states (corridor endpoints)—where rewards are encountered during navigation—are learned using a separate CSCG. The cognitive map derived from this process encodes the sequences of goals the simulated creature must attain to secure rewards (as depicted in the Level 2 box of Figure 6B). It is important to note that this map aligns with task space rather than physical space. For instance, both nodes labeled "C" correspond to the same physical location (the center corridor), but signify distinct task stages: specifically, reaching the center after either visiting the left or right corridors.

Following the learning phase, the ensuing generative model enables the spatial alternation task depicted in Figure 6A to be solved. This is achieved through a combination of hierarchical perceptual inference for self-localization and hierarchical planning to determine the subsequent course of action. During perceptual inference, the model combines its existing beliefs ("from the past") with its current observations ("from the present") to infer its position within both the cognitive and task-space maps. The incorporation of message passing "from the past" provides the system with a working memory of previous contexts—an essential attribute for successfully addressing spatial alternation tasks [39,40].

During the hierarchical planning phase, the model makes choices—regarding the upcoming states to visit—in both the task and physical spaces to attain rewards. The process starts at the higher level, where the model considers plans for the next state in the task space map (in essence, determining which corridor to approach in alignment with the rule). Subsequently, this goal is set as a prior within the lower-level model. Then, the lower-level model engages in spatial planning, selecting a sequence of actions within the physical space map. Despite utilizing distinct maps at the two levels, the planning procedure remains internally consistent. It involves selecting a sequence of actions (referred to more formally as control states) anticipated to minimize expected free energy ($G(\pi)$), as depicted in Figure 6D.

Expected free energy—crucial to the planning process—is distinct from the variational free energy introduced in Figure 2. Unlike variational free energy, which considers only past and present observations, expected free energy factors in future observations, expected under each policy. The ability to forecast future, policy-dependent (and counterfactual) observations enables the evaluation of policies based on two criteria outlined in Figure 6D: "epistemic value" (or expected "information gain") and "pragmatic value" (or expected value). The former criterion gauges the extent to which policy-dependent observations could modify the organism's posterior beliefs, favoring policies that offer informative observations. While navigating the W-maze shown in Figure 6A, this emphasizes policies leading to unambiguous states, like the bottom-center of the maze where observation 16 is available. Conversely, pragmatic or instrumental value assesses how effectively policy-dependent observations align with the organism's preferred observations (represented as a prior over observations, indicated as C), prioritizing policies that achieve prior preferences. In the context of our spatial navigation illustration, this implies favoring policies leading to rewarding states; i.e., plans that realize the prior belief that securing rewards is characteristic of the agent question.

The simulations presented in [38] offer a compelling demonstration of the effectiveness of hierarchical planning in solving the spatial alternation task, as depicted in Figure 6E. Moreover, experimental evidence from studies disrupting hippocampal activity (Jadhav et al., 2012) concurs with these simulations, revealing that performance deteriorates at challenging decision points (specifically, within the center corridor), when communication between the two hierarchical levels of the model is interrupted. This breakdown occurs because communication interruption prevents distinguishing between the two "C" states. The same model is also capable of mastering more intricate tasks, necessitating rule switches intermittently. Notably, the model accurately replicates the empirical finding that interrupting communication between the two hierarchical levels hampers rule switching and promotes perseverative behavior, in accordance with studies conducted by [44].

*Generative replay*

In addition to planning, generative dynamics can support other functions. Substantial evidence indicates that during periods of inactivity, distinct brain regions—including the hippocampus, prefrontal cortex, ventral striatum, and visual cortex—engage in "replay" revisiting sequences of experienced content like past navigation trajectories. Importantly, this replay process also involves the generalization and novel recombination of encountered content [45–54]. The phenomenon of replay activity, along with the inherent activation of neuronal assemblies, can be explained as the intrinsic dynamics of a generative model. This dynamic activity serves to optimize the model during periods of rest, such as when novel observations are absent. This mechanism allows the brain to optimize its generative model for future use, even in the absence of external input [55,56].

A further illustrative instance of hierarchical generative modelling emerges from a study that addressed the formation of spatial maps using a probabilistic mixture model as depicted in Figure 7A [57]. While this model is similar to the model showcased in Figure 6, it explicitly encodes maps and sequences. At the highest level, the model acquires various spatial maps corresponding to distinct mazes or segments within a single maze, alongside an associated probability distribution allowing to identify the currently most plausible map. The lower levels capture sequences of spatial positions and individual spatial locations. In this framework, self-localization involves inferring the current map, sequence within the map, and position in the sequence. This inference task entails the fusion of top-down and bottom-up information. For instance, determining the current location relies on integrating sensory observations from below and the top-down information derived from the inferred sequence and map. This process utilizes a message-passing structure akin to that found in hierarchical predictive coding networks.

When subjected to a continuous learning task that involves acquiring multiple spatial maps (as illustrated in Figure 7B), the model developed distinct maps for individual mazes, as depicted in Figure 7C. Notably, the model's performance in the ongoing learning demonstrates improvement when employing a technique known as "generative replay." This entails generating fictitious sequences of observations by sampling from the learned maps at the highest level and propagating this information down to the lower layers. These replayed sequences are then utilized to facilitate the self-training of the model. Given that the maps encode probability distributions over locations, and significant sites like goal locations and bottlenecks are assigned higher probabilities, generative replay naturally favors these prominent positions: a tendency confirmed through empirical observations. Moreover, it is important to note that the replayed sequences are not mere replicas of experienced sequences but can recombine them in novel ways, as evidenced by prior studies [47]. The outcomes presented in Figure 7D reveal that the deployment of generative replay not only enhances map learning but also allows the model to outperform baseline models that do not employ replay, as well as models that engage in self-training through experiences stored in external "verbatim" memory sources (referred to as experience replay). These findings highlight the efficacy of generative replay—and the offline reactivation of content originating from the generative model—in optimizing the model even when no new observations are available. It is worth noting that this same technique of generative replay could potentially find application in training other models, such as semantic memories [58] or action controllers [59].

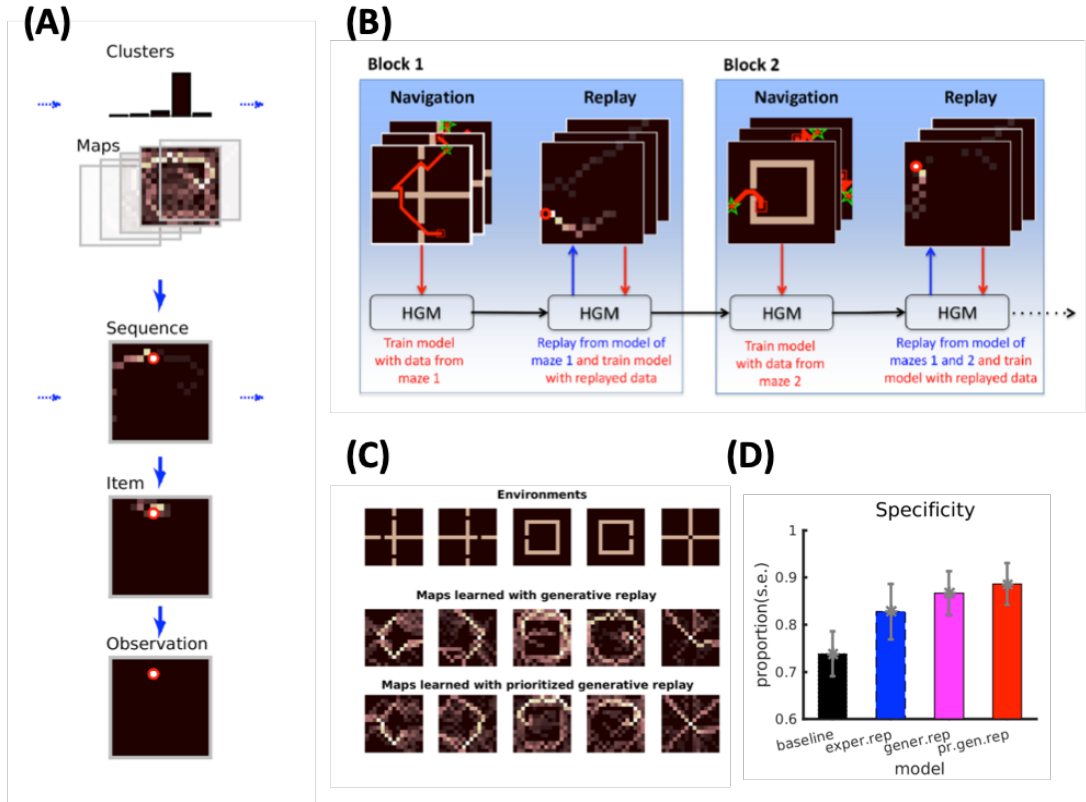

*Figure 7. Hierarchical Generative Model for Spatiotemporal Sequences, adapted from* [57]. *(A) The hierarchical generative model encompasses three distinct sets of hidden states: maps, sequences, and items (e.g., spatial locations), to explain sensory observations. (B) The training process alternates between real-life navigation experiences and "generative replay". (C) This strategy facilitates continuous learning by acquiring multiple maps, thereby mitigating catastrophic forgetting. The colors denoting learned maps denote probability distributions across locations, peaking at salient points like bottlenecks or goal locations. (D) Comparative analysis of different versions of the same hierarchical model, encompassing baseline without replay, experience replay (exper. rep), and two forms of generative replay (gener. rep and pr. gen. rep), demonstrates the efficacy of generative replay in enhancing map learning. See* [57] *for details.*

*Generative models, belief dynamics, and neural representation*

The generative models featured in this section offer valuable insights into the concept of neural representation. These examples help differentiate between two potential interpretations of neural representation. First, generative models might encode "cognitive maps" that pertain to physical space or task space. These maps could have their parameters encoded in synaptic weights that change gradually with learning over time. Second, the (approximate) posterior beliefs, whose parameters could be encoded by changing neural activity within neuron assemblies, could stand for the organism's current best assessment of its position within these cognitive maps.

It is crucial to recognize that the "representational content" of these beliefs hinges on the variables inherent to the generative model. For instance, in the hierarchical generative model in Figure 6, the system accomplishes self-localization in both physical and task space. This means that it maintains probabilistic beliefs regarding its present location within the maze at the lower level, and simultaneously keeps track of the current stage within the spatial alternation task at the higher level. On the other hand, the model depicted in Figure 7 offers a slightly different perspective: it implies the representation of a broader spatiotemporal context, encompassing the ongoing sequence and map. This effectively combines sensations, predictions, and memories. Of course, it is important to note that both of these models are considerably simplified in comparison to the intricate operations of hippocampal, prefrontal, and other brain circuits. For related insights into hippocampal dynamics, including factorized representations of "what" and "when", interested readers are referred to [60]

Importantly, both planning and generative replay extend beyond mere self-localization, giving rise to beliefs and observations that potentially encompass the present, past, and future. Beliefs regarding the future cannot be directly inferred from current observations; instead, they necessitate the engagement of the generative model to internally generate (or imagine) future observations. These internally generated dynamics showcase forms of cognition that are "detached" from the immediate context, going beyond the constraints of the present moment (c.f., the specious present of James and Husserl). These types of cognitive processes align with the conventional sense of representation in cognitive science: the capacity to represent objects or events even *in their absence* [61]. The ability to represent things in their absence is especially valuable for advanced cognitive functions that involve foresight and planning for the future. This concept could be instrumental in understanding brain reactivations and replay phenomena. For instance, one could speculate that generative dynamics occurring just before navigation resemble a form of "local replay," optimizing the next plan, while generative dynamics during periods of rest could be likened to "remote replay," optimizing the model for future interactions [62–64].

In terms of evolutionary advantage, the capacity of generative models to "detach" from immediate circumstances could have conferred substantial benefits [65]. This ability to engage in internally generated simulations and predictions, thereby envisioning scenarios beyond the immediate context, might have played a significant role in the adaptive success of organisms throughout evolution. Active inference—an application of the free energy principle—licenses this kind of teleological (or perhaps teleonomic) account; in the sense that to minimize variational free energy is to maximize the marginal likelihood of sensory exchanges with the eco-niche, which scores the adaptive success of a phenotype.

Lastly, a brief consideration on how generative models acquire "meaning" and the ability to represent something is useful. The two aforementioned studies on cognitive map formation depict generative models as *templates* or predefined frameworks of maps or schemas [57,66]. Initially, these templates lack inherent meaning, serving as vessels for encoding experiences and generating predictions. The acquisition of meaning happens when these templates accumulate sensorimotor experiences. In essence, these generative models transition from being meaningless to acquiring meaning, such as signifying specific spatial locations or positions within task spaces. This transition occurs as they become linked to the external world through action: a process that [67] aptly describes as the journey from the realm of the meaningless to that of meaning through interaction with the environment.

**Abstraction and distortion in generative models and beliefs**

Until now, our focus has centered on generative models of the external world, particularly models read as cognitive maps. However, it is important to acknowledge that models are never

"perfect replicas" of reality. A principled reason for this divergence lies in the principles of active inference, sometimes referred to as self evidencing [68]: to minimize free energy is to maximize marginal likelihood or model evidence. Crucially, log evidence can always be decomposed into accuracy and complexity. This means that generative model learning achieves a balance between *accuracy*—encompassing data reconstruction and prediction—and *complexity*, considering factors like the number of model parameters or degrees of freedom. The process of penalizing complexity results in generative models whose hidden states incorporate only relevant details that enhance accuracy.

In essence, the most effective generative model tends to retain sufficient detail to accurately represent observations and achieve task success (in the case of action-guiding models), while discarding redundant information that does not bring an increase in accuracy. Consequently, learning leads to the development of parsimonious models that abstract away from irrelevant details, forming compressed latent spaces or manifolds [55,69–71]. This is evident in the cognitive maps discussed earlier. However, it is worth noting that the implicit geometry can diverge from the true geometry of the external world. Such distortions are also present in the brain's generative models. For instance, studies have indicated that hippocampal cognitive maps are not uniformly distributed across the environment. Instead, a greater number of place cells are typically found in locations tied to goals and other significant features [72–74]. Moreover, spatial locations with equal value tend to cluster more closely in the latent or hidden state space, further exemplifying these distortions [70].

One plausible computational explanation for these distortions is found within the realm of *model reduction*: a process applied after a model is learned to fine-tune the tradeoff between accuracy and complexity. After model learning, model reduction assesses whether each model parameter (c.f., synaptic connections) contributes a commensurate level of accuracy to outweigh its complexity costs. Parameters that do not meet this criterion are pruned, effectively removing them from the model's architecture (and by pruning synapses) [55]. This practice inevitably entails the potential loss of certain information, especially concerning infrequent observations. However, it simultaneously results in more efficient models that offer improved generalization by avoiding overfitting.

Rate distortion theory [75] and algorithmic complexity [76,77] furnish convergent perspectives on the interplay between accuracy and complexity during model learning. See also [78–82]. Unlike model reduction, which tends to be a post-learning procedure, rate distortion theory operates during learning itself. It formalizes how a model with a fixed capacity—with finite degrees of freedom—can optimally compress information, mirroring the statistical distribution of observations and their functional relevance. This process engenders systematic distortions in the representation, whereby frequently encountered observations and those most pertinent to a particular task are encoded with higher fidelity, while less crucial observations are compressed to a greater extent.

To illustrate this concept, let's consider a generative model commonly used for dimensionality reduction in machine learning; namely, the β-variational autoencoder [83]. We will use this model to learn a set of 500,000 "spatial maps," which are 13x13 black and white images containing two vertically oriented, noisy "corridors." These corridors are positioned either in the upper or lower portion of the image. In Figure 8A, we present 21 example maps, each labeled based on the pixel positions of the centers of the two corridors. For instance, the label (0,12) indicates that the center of the upper corridor corresponds to pixel 0, while the center of the lower corridor aligns with pixel 12. Our analysis involves a comparison of four different versions of the β-variational autoencoder, each possessing distinct characteristics. The first two models (Model 1 and Model 2) are β-variational autoencoders with standard unsupervised learning objectives. These objectives have the functional form of variational free energy but include a hyperprior β, called *capacity* that scales the complexity (i.e., rate) in relation to accuracy (i.e., distortion). The difference between these models lies in their capacity: Model 1 has a high capacity, while Model 2 has a low capacity (as depicted in Figure 8B). The remaining models (Models 3 and 4) are also β-variational autoencoders, but they incorporate supervised learning. In particular, this second objective pertains to the accuracy of binary classification. The task is to recognize whether the input image features aligned corridors (classified as class 0) or non-aligned corridors (classified as class 1). Although the learning objectives are consistent across Models 3 and 4, they diverge due to their capacity: Model 3 possesses high capacity, whereas Model 4 has low capacity.

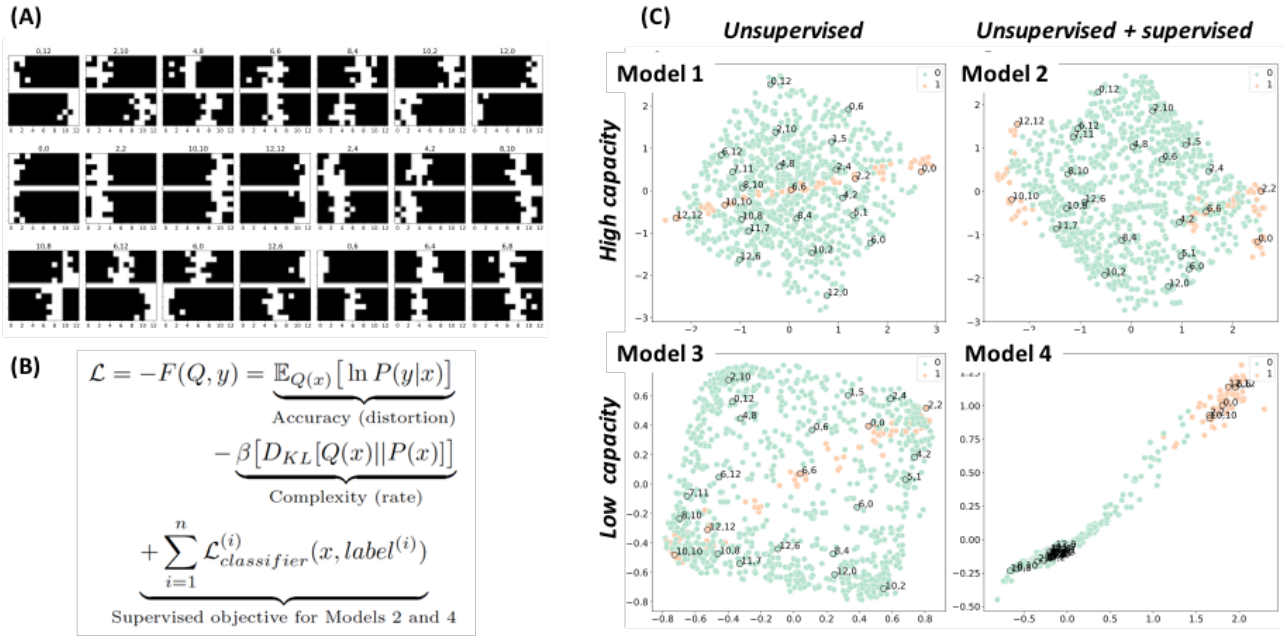

*Figure 8. Abstraction and Distortion in Generative Models. This illustration presents latent codes acquired through β-variational autoencoders (Higgins et al., 2016), designed to encode and compress spatial map images. (A) Depicts 21 samples from 500,000 spatial maps utilized for learning, featuring two noisy vertical "corridors" placed in the upper and lower image portions. Map labels signify the upper and left corridor centers spanning 13 pixels (0 to 12). (B) Details the employed objective (loss) functions during training. The first two lines delineate terms for unsupervised learning in Models 1 and 3. These correspond to a (negative) free energy, which is decomposed here into accuracy and complexity terms (not into evidence and divergence as in Figure 2). These two terms align with the distortion (i.e., accuracy) and rate (i.e., complexity) facets of rate-distortion theory* [75]. *The parameter β in the second line adjusts model capacity; Models 1 and 2 have higher capacity, while Models 3 and 4 have lower capacity. The third line pertains to supervised (classification accuracy) terms used solely by Models 2 and 4. This supervised objective involves binary classification, distinguishing alignment (class 0) or misalignment (class 1) of the two corridors in the input image. (C) Charts latent codes generated by the four models. Each dot corresponds to one image's position in the generative model's low-dimensional latent space; certain images are labeled (e.g., 6,6*

*at the center) for clarity. Dot colors reflect image class in the classification task (green for class 1, orange for class 0).*

The four panels of Figure 8C offer a visualization of the 2D projections of latent codes learned by the four distinct β-variational autoencoders. Each dot corresponds to the position of a specific example image within the low-dimensional latent space of the generative model. For clarity, certain images are labeled with positions, such as "6,6" in the center. Additionally, the color of the dots indicates the class to which the image belongs in the classification task: green denotes class 1, while orange signifies class 0. In the case of Model 1 (top-left panel of Figure 8B), the latent codes expand across the factors of variation (namely, the positions of the upper and lower corridors) in an orthogonal manner. Model 2 (bottom-left panel of Figure 8B) demonstrates similar latent codes, yet certain regions are more densely clustered, resulting in a loss of resolution. Models 3 and 4 exhibit latent codes wherein inputs belonging to class 0 are pushed towards the periphery, leading to improved discrimination. This phenomenon is particularly pronounced in Model 4, where the latent codes are binarized, causing them to span the factors of variation unequally. These examples underscore the influence of balancing *accuracy* (i.e., distortion) against *complexity* (i.e., rate). In active inference, the capacity *per se* is optimized in terms of precisions that are inferred in the usual way (see below).

To summarize, the above numerical studies show that generative model learning favors low-dimensional latent spaces or manifolds [69]. These latent state spaces or manifolds are shaped by the statistical characteristics of the observations and, in the case of task-driven models, the nature of the task itself. However, the trade-off between accuracy and complexity implies that these latent spaces are distorted in an optimal way. For instance, latent spaces might emphasize the most frequent inputs. Moreover, this implies that within living organisms, where observations are contingent upon actions—and not predetermined as in typical machine learning scenarios—the learning history plays a role in shaping the geometry of latent spaces.

*Distortions in beliefs and precision control*

Not only the structure of generative models but also Bayesian beliefs can exhibit distortions. Interestingly, this can occur even when there's a direct one-to-one correspondence between the internal hidden state of the generative model and the external hidden state of the world. For instance, consider the case of the "finger angle" discussed in Figure 3, where the angle is the defining characteristic. Even in this scenario, an organism's posterior belief regarding the finger angle might deviate from the true angle. This divergence can stem from various factors. For example, the inference process might converge toward an incorrect value due to noisy or misleading observations. This situation is akin to perceptual illusions or instances of "misrepresentation."

A principled account of this kind of optimal "misrepresentation" is that, in the context of active inference—and Bayesian inference more broadly—the outcomes of inference are determined by the interplay of two significant factors: observations and priors. In active inference, these priors represent an organism's preferred states, leading to the possibility that an agent's posterior beliefs might not accurately represent the external reality. Instead, these beliefs could be biased due to the incorporation of preferences, giving rise to a form of "optimism bias." For a deeper discussion on this concept in relation to representation, the reader can refer to [13]. This concept is demonstrated in Figure 3C, which highlights how the inference process concerning angle position tends to converge towards a midpoint between the true finger position and the prior when actions are precluded. This example underscores a fundamental distinction between active inference and other frameworks, wherein the relationship between perceptual inference and action is not considered.

The simulation presented in Figure 3C assumes that priors and observations carry equal weight, but in reality, this isn't a general rule. In predictive coding and active inference frameworks, the weighting of information sources is determined based on their *precision* (or inverse variance, in the case of a Gaussian distribution). This means that information with higher precision, indicating lower uncertainty, is given more weight and consequently has a stronger influence on the inference process. This precision-weighting mechanism plays a crucial role in both perceptual inference and movement control. Incorrect assignment of precision values can lead to disorders in both domains. For instance, some disorders associated with schizophrenia, such as hallucinations and persistent distortions of perception, might result from overly precise priors, causing the inference process to disregard sensory observations and rely excessively on internal models [84].. This mechanism can extend to various other psychopathological conditions and motor disorders where the misalignment between precision settings of priors and sensory inputs can lead to impairments [24,26,85–90]. Collectively, these studies emphasize the importance of precision-weighting within generative models, as it plays a critical role in fine-tuning an organism's posterior beliefs and their alignment with (largely self generated) reality.

**Action-oriented generative models: neural representation with an enactive flavor**

*"In a true sense, for example, the frog does not detect flies—it detects small, moving, black spots of about the right size. Similarly, the housefly does not really represent the visual world about it—it merely computes a couple of parameters…which it inserts into a fast torque generator and which cause it to chase its mate with sufficiently frequent success."*
– David Marr, 1982.

So far, we have considered a rather classical concept of representation, in which the organism's generative model plays the role of a "small scale" model or cognitive map of something out there (for example, a maze) and neural activity represents some entity in the map (e.g., one's current location or a future predicted location in the map). While a cognitive map does not include every detail and can be distorted, it still features a correspondence between hidden variables in the brain (e.g., locations in a cognitive map) and in the external world (e.g., locations in a maze). This is a standard assumption in "representational" cognitive science, which assumes that models encode hidden variables isomorphic to things like objects or places in the external world, in order to infer them.

However, active inference introduces a novel angle, suggesting that the brain's generative models are primarily developed to enable adaptive action in the world, rather than just understanding it. This aligns with the "good regulator" theorem, which asserts that an effective regulator of a system needs to possess a model of that system [91]. In this context, the hidden variables within action-controlling models do not necessarily mirror the hidden variables of the external world. In fact, the key purpose is to guide appropriate actions, with precise representation being secondary. This means that the level of detail or accuracy in how a generative model corresponds to external reality becomes less crucial, if the model is effective in guiding adaptive action. As long as the model supports successful decision-making and response to changes in the environment, the extent to which it faithfully represents the external world becomes less important within the framework of active inference. This perspective shift underscores the functional and pragmatic aspects of representation, where accuracy in mirroring external reality is not a strict requirement as long as the model serves its purpose in facilitating adaptive behavior.

Acknowledging the pragmatic function of generative models in facilitating interactions opens up *enactive* interpretations of active inference, which prioritize adaptive action above mere representation and occasionally even challenge the concept of internal representation. An intriguing aspect is that, in contrast to traditional philosophical dichotomies between classical cognitive theories and enactive theories, active inference does not mandate the exclusion of one for the other. This is due to the fact that active inference reconciles the pursuit of *interacting with* and *representing* the world. Consequently, a pertinent query emerges: to what extent should an organism's generative model mirror the external world for optimal efficacy?

In a range of tasks, a spectrum of generative models are effective, spanning from "explicit" or "environmental" models – exemplified by cognitive map models in Figures 6-7 – to "action-oriented" or "sensorimotor" models, which prioritize interaction, while sidestepping the encoding of external world variables [12,14,92,93]. It is important to bear in mind that generative models encode the underlying causes of observations. The key distinction between explicit and action-oriented generative models lies in their explanations for observable sensations: the former attribute these sensations to the consequences of external states of affairs (e.g., an image on the retina arising from an external object), while the latter attribute them to the outcomes of actions (e.g., touch sensations stemming from whisking movements).

Despite their differences, both types of generative models can effectively support the same tasks. For instance, consider the scenario where a rodent needs to gauge the distance from an object. It has the option of employing an explicit generative model, which provides detailed descriptions of external elements; encompassing object identities and spatial distances. This model enables distance (and object) estimation by inferring the most probable values of its hidden states, akin to how the finger control model infers finger angles or the spatial navigation model infers the animal's likely location.

On the other hand, the rodent has the option of utilizing an action-oriented model that exclusively captures relationships between whisker movements and touch sensations, without explicitly incorporating variables representing external entities like objects. A prime illustration of this concept is the active whisking model presented by [94], which replicates empirical observations of anticipatory whisker control [95], as depicted in Figure 9. Within this generative model, the magnitude of whisker oscillation depends on a hidden state, which combines contributions from both a fixed central pattern generator and a prior about the desired amplitude. Significantly, this model predicts a touch sensation when the whisker is protracted, with both prior beliefs and whisker dynamics governed by the minimization of prediction errors until this prediction is fulfilled. In cases where the anticipated touch sensation at the protraction's conclusion does not materialize, the prior is adjusted, resulting in an increased whisker amplitude in subsequent cycles. Conversely, an unpredicted touch sensation leads to an update of the prior in the opposite direction, resulting in a reduced whisker amplitude in the ensuing cycles. Upon convergence and subsequent minimization of prediction error, the animal-object distance can be read out from the hidden state encoding whisker amplitude, despite the model not inherently encoding distance: c.f., perceptual control theory and related formulations [96,97].

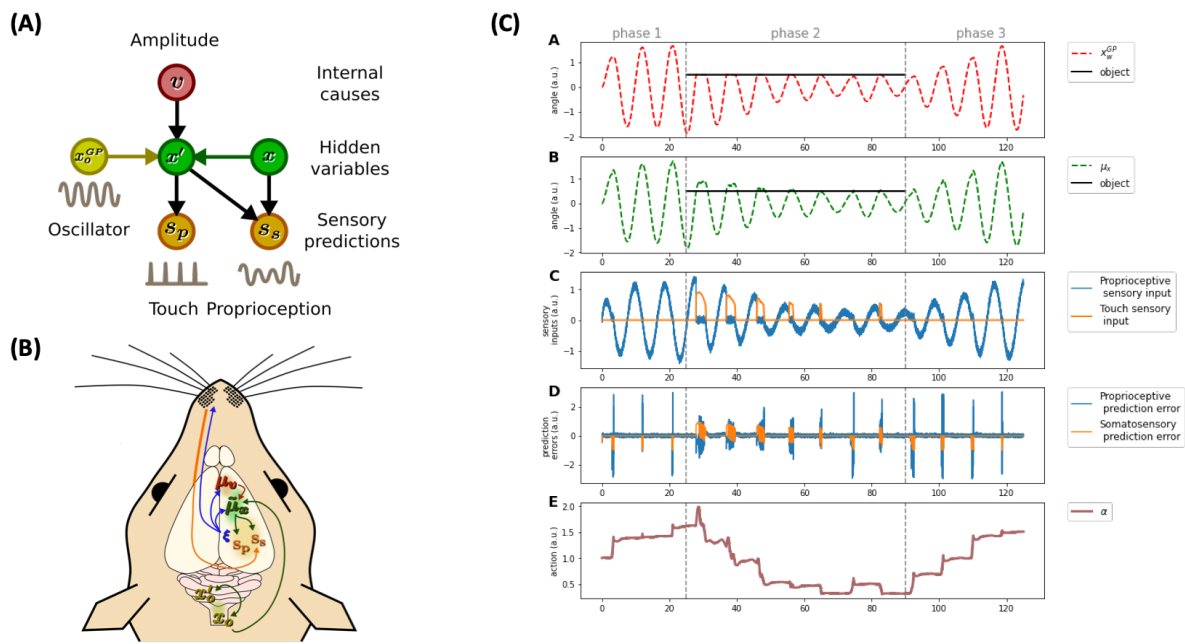

*Figure 9. Active Inference Model of Whisker Movements, adapted from [94]. (A) Displays a schematic representation of the generative model with a prior over the desired amplitude, hidden variables governing whisker amplitude and generating somatosensory (touch) and proprioceptive predictions. (B) Depicts the temporal evolution of several model variables when the whisker unexpectedly contacts an object (horizontal black line) and when the object disappears. In Panel A, the actual whisker amplitude is shown, halting upon encountering an object. Panel B illustrates the inferred amplitude; during initial contact, it surpasses the object but swiftly converges to the object's distance in subsequent cycles. Panels C and D illustrate proprioceptive observations and prediction errors, respectively. Panel E showcases action dynamics, increasing amplitude prior to touching an object and reducing it after unexpected contact.*

This example underscores the potential to regulate actions using action-oriented generative models, which encapsulate sensorimotor contingencies, occasionally being characterized as non-representational [98,99]. Other action-oriented generative models can also be employed to implement diverse embodied strategies. For instance, strategies that enable rodents to navigate a maze without constructing a representation of it, instead altering their course when encountering obstacles [10]. Similarly, strategies could be devised for a baseball outfielder to catch a ball by choosing a running path that counteracts the optical acceleration of the ball, without necessitating the prediction of its trajectory [12,100–103].

Another instance of an action-oriented generative model can be observed in the nervous system of a very simple organism: Caenorhabditis elegans. Multiple studies have demonstrated that a significant portion of the brain activity of C. elegans encompasses sequences of locomotion behaviors, including forward and reverse locomotion, as well as ventral turns, which scaffold the animal's action selection [104]. Dimensionality reduction analyses suggest that these behavioral sequences are arranged into stable cycles. The points at which these cycles bifurcate might correspond to decisions between alternative behavioral sequences. For instance, a decision point could involve choosing between forward locomotion and reverse locomotion after executing a ventral turn, as depicted in Figure 10, taken from [105]. These findings indicate that a substantial portion of C. elegans' brain activity is structured to facilitate sequential interactions with the external environment, without explicitly encoding the external world's state. This generative model proves sufficient for C. elegans to navigate its surroundings without building an internal representation of it. Similarly, this model could potentially utilize feedback sensations to make adaptive locomotion decisions at bifurcation points, for example, increasing the likelihood of a reverse movement after a ventral turn if an object is detected in the front, or decreasing it if an object is sensed from behind. This mechanism would conceptually resemble the adjustment of whisker amplitude following an unexpected touch sensation (or an increase following a lack of expected touch sensation) as observed in the active whisking model proposed by [94].

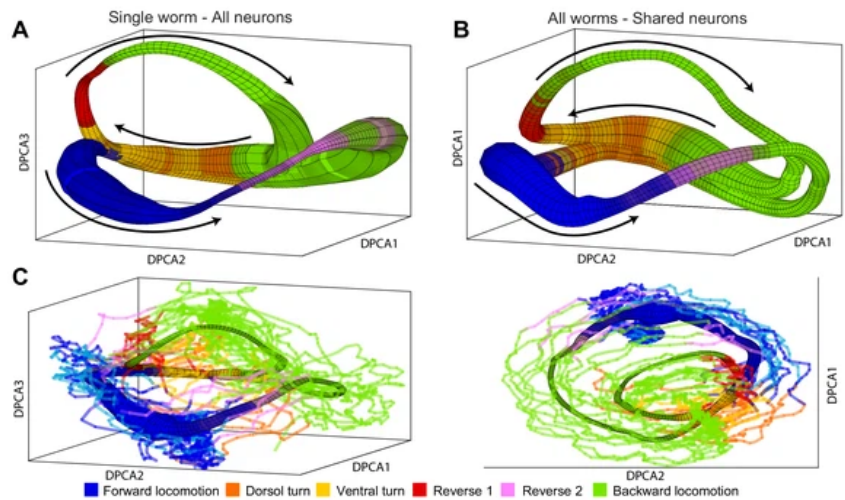

*Figure 10. Neural Population Activity Representation in Caenorhabditis elegans, adapted from* [105]*. A low-dimensional depiction of neural population activity in the Caenorhabditis elegans, showcasing cycles of locomotion behaviors, color-coded to represent forward locomotion, reverse locomotion, ventral turns, and more. The diagram captures bifurcation points that could signify decisions between distinct behavioral patterns. See* [105] *for details.*

In summary, both extrinsic and action-oriented models, or even hybrid models combining aspects of both, have the potential to support cognitive tasks. However, only the extrinsic models would additionally involve the reconstruction of the external environment. If reconstructing the external environment aligns with the traditional notion of representation, then it becomes an empirical inquiry to determine which species or tasks entail a neural representation in the conventional sense. This leads to two methodological considerations. Firstly, as previously discussed, identifying neural representations in an organism means disclosing the generative model it employs for making inferences. Secondly, the mere ability to read out certain information, such as distance details, from an organism's behavioral or neural variable does not automatically indicate that the organism is representing this information, at least not in the classical understanding of representation. A diagnostic query arises: whether the distance information is utilized by the same neural area or downstream circuits, for instance, to make decisions.

**Discussion**

In this article, we have explored the concept of neural representation through the lens of active inference, a normative framework that offers insights into brain function and cognition [10]. We conclude by reviewing five key discussion points concerning the mechanisms of active inference. We then underscore the implications of these points for the concept of neural representation.

First, a fundamental imperative, guiding living organisms, is the minimization of the discrepancy between their predictions and observations. This imperative is realized through action and perception, which jointly minimize *variational free energy* (or prediction error under certain conditions). Second, the generation of predictions and requisite inferences hinges on the acquisition and use of generative models; namely, statistical models that describe how observations are generated from unobservable (hidden or latent) causes, e.g., how a visual object, such as an apple, generates an image on the retina. *Generative models* support predictive and inferential processes both within perception (predictive coding) and action (active inference). Third, more advanced generative models afford planning, imagination, and forward-thinking cognition. Planning necessitates the utilization of the generative model to envision future observations [106–109], enabling the assessment of potential action plans by gauging their capacity to minimize *expected free energy*. Consequently, planning embodies a form of cognition *detached* from the present moment. Fourth, the acquisition of generative models balances accuracy and complexity. Optimal generative models embody a trade-off between these constituents of model evidence, fostering abstraction while introducing apparent "distortions" relative to reality. Fifth, within the framework of active inference, generative models serve the purpose of enabling adaptive action. While providing a comprehensive or even veridical depiction of the external world might prove beneficial on occasion, it is not an absolute necessity. This implies that various organisms can effectively employ diverse types of generative models. These range from "explicit models" encompassing variables for entities such as objects, faces, or individuals, to "action-oriented models" that facilitate adaptive action solely by predicting the consequences of actions without encoding external entities like objects. Each of these points has implications for the notion of neural representation.

### *Free energy minimization operates at multiple timescales*

When highlighting the significance of free energy minimization, it is important to acknowledge that this process operates across multiple temporal scales within the brain's mechanisms (e.g., rapid neural firing rates versus gradual synaptic updates) and is intertwined with action, which unfolds over diverse timescales (e.g., basic movements versus complex action sequences aimed at achieving distant goals) [110]. During the action-perception loop, the minimization of free energy changes the organism's beliefs about the world and guides actions to shape the world. This pertains to the timescale of neuronal firing rates. On the other hand, learning mechanisms and the adjustment of model parameters operate at a slower timescale relative to the action-perception loop, manifesting as synaptic updates. Moreover, learning processes also manifest during periods of rest, during which the brain potentially optimizes the generative model for future use via model reduction and synaptic pruning [55,56]. The optimization of generative models also extends to the slower temporal scales of development and evolution. One can conceivably interpret these gradual adaptations of generative models to an organism's ecological niche as a form of very gradual Bayesian model selection [111–113].

Conversely, planning and imagination might operate at a faster timescale relative to the action-perception cycle. In the context of planning, an organism can refine its beliefs concerning future actions and the latent states it will encounter (and retrospectively modify beliefs about the past). This domain is characterized by the timescale of "replay" events. Lastly, the process of free energy minimization encompasses modifications that extend beyond the confines of the brain, such as joint action and cultural dynamics [114–118]. However, these aspects are not covered within the current discussion.

Recognizing the diverse timescales at which free energy minimization operates is pivotal for contextualizing neural representation within a broader framework that extends beyond the immediate operations of the brain. As far as neural representation is linked with beliefs and generative models (as detailed below), it's noteworthy that these elements are optimized through intricate nested processes, unfolding at varying temporal scales, from neuronal activity through to evolution. This perspective is important to *naturalize* the notion of neural representation: namely, recognize that it might be an inherent adaptation that empowers organisms to minimize their free energy—or to maximize the evidence for their existence [10,11].

### *Generative models, belief dynamics, and neural representation*

The concepts of a generative model and belief dynamics are inherently intertwined with the notion of representation, since the inception of cognitive science [17,18]. However, the idea of representation can be approached in at least two distinct ways. On one hand, the generative model itself, whose parameters could be encoded within synaptic weights, could align with the concept of neural representation. This interpretation is akin to referring to the "representation" of a "cognitive map." Conversely, the dynamics of beliefs, which might correspond to neural firing rates, could also be conceived as "neural representations" of the subject matter of those beliefs—such as the position of a finger or the multimodal concept of a leaping frog. The latter perspective is more closely associated with the traditional attributes of (neural) representation: they are *about* something and are employed by the organism *to accomplish specific tasks*, like predicting the sensory outcomes of actions or guiding adaptive behaviors. Consequently, they bear meaning and significance for the organisms themselves [1,3,5–8].

### *Planning, imagination and detachment*

A fundamental distinction exists between variational free energy, foundational to action and perception, and expected free energy, which forms the basis of planning. The former solely relies on past and present observations, while the latter necessitates the utilization of the generative model to endogenously generate (or imagine) counterfactual observations that could arise from a particular sequence of actions. This mechanism has implications not only for planning but also for various manifestations of future-oriented cognition, encompassing concepts like "as-if" simulation, re-enactment, and imagination; as well as the re-enactment of past episodes or the construction of counterfactual events. It's worth highlighting the term "endogenous," as this generative capacity hinges on temporarily detaching from the immediate action-perception loop, thereby disregarding sensory inputs and prediction errors.

From a neurological standpoint, the ability to detach and replay statistical patterns encoded within the generative model, revisiting past experiences or envisioning future scenarios, finds clear parallels in neuronal reactivations and replay [45,46]. Theoretically, this phenomenon aligns with the classic notion that representation serves to represent something "in its absence" [61]. Nevertheless, not all agents engaged in active inference partake in planning or expected free energy minimization. Thus, if the capacity for detachment and the representation of something "in its absence" differentiates between organisms that possess or lack representation (or cognition), it potentially underscores a categorical division between active inference agents exclusively minimizing variational free energy and those additionally minimizing expected free energy [119]. Consequently, inquiries about an organism's capacity to represent, the manner in which it does so—and whether it engages in such activity—again rests on the identification of its underlying generative model.

*Abstraction and distortion in generative models*

Even when adhering to the conventional definition of a model as a representation of the external world, it is important to acknowledge that no model can perfectly mirror reality, and some degree of information loss is inevitable. However, paradigms such as active inference and rate distortion theory [75] go beyond this obvious consideration and underscore the significance of optimal information compression within generative models. Within the active inference framework, generative models must strike a balance between accuracy and complexity. This balance aims to facilitate optimal learning from data, ensuring both efficiency and the ability to generalize. As depicted in Figure 8, distinct constraints on model learning—such as variations in data statistics and learning objectives—lead to diverse geometries and alterations in the latent states that generative models acquire.

Recent research into the geometries of neuronal responses across various brain regions has gained momentum [69,70,120]. To the extent that these geometries reflect how we perceive the world, it becomes evident that not only the generative model itself but also the specifics of its learning statistics play a crucial role in determining what it represents and how it does so.

*Explicit and action-oriented generative models*

The brain learns generative models to navigate the world adaptively, not (or not solely) to understand it. Different living organisms may possess an array of generative models, spanning from "explicit" models that involve variables for inferring external entities like objects, faces, or people, to "action-oriented models" that prioritize predicting action outcomes, omitting the need to encode external entities. These models can offer varied solutions to shared biological challenges, such as estimating distance from an object. Therefore, the specific model an organism employs for a particular task remains a matter of empirical investigation.

This perspective licenses "enactive" or "pragmatic" viewpoints concerning the role of generative models, which emphasize that the role of generative models whose main role is mediating interaction and not representing the external world veridically. From a theoretical stance, this notion can be interpreted in several ways. One interpretation could suggest that action-oriented generative models still facilitate neural representations, though these might be understood as "pragmatic representations" conveying affordances and potential actions. Another interpretation might assert that models of sensorimotor contingencies do not necessitate the concept of representation [99]. While resolving this debate is beyond our scope, one can note that diverse generative models (explicit or action-focused, with varying hidden variables) equip living organisms with distinct capabilities. For instance, as previously discussed, generative models in active inference enable adaptive action-perception loops, but only certain models support detached cognitive operations such as planning or imagination.

This observation has a significant consequence: assuming that "meaning" and "understanding" are inherently connected to agentive action through generative models, the array of generative models described above imparts diverse forms of "understanding" of an organism's ecological niche. This spans from the "meaning-in-interaction" foregrounded by enactive theories to the "representation-in-the-absence-of-the-reference" accentuated by cognitive theories. Similarly, generative AI models that lack engagement in agentive interactions with their surroundings would lack the essential grounding in reality necessary for fostering genuine understanding (Pezzulo et al., 2023).

***The symbol detachment problem: from pragmatic to descriptive representations***

The preceding discourse raises a crucial, yet unresolved query: how did the relatively simpler generative models, enabling action control in our early evolutionary ancestors (as exemplified by C. elegans in Figure 10), evolve into those that also facilitate detached cognition? Various researchers [4,65,121–125] have proposed a plausible transition from "pragmatic" to more "detached" and "descriptive" representations, which can be condensed into two pivotal steps.

Initially, meaningful activity patterns could emerge within the context of an organism's capacity to predict and control simple behavioral strategies. These could be deemed "pragmatic" neural representations if they are *about* something: for example, they provide predictions about sensory feedback ensuing from object interactions, albeit only in restricted contexts. These patterns are more than mere correlations—they exhibit a systematic and "exploitable relation" with external elements (whether explicitly encoded or not encoded within the model's hidden variables). This enables them to *guide adaptive actions*, imbuing them with meaning and adaptive significance. These representations are *grounded* in the organism's capacity for adaptive prediction and control, but they are not detached: they are context-bound, and only activated in the context of—and proximally coupled to—the action-perception loop and the current state of the organism. An instance highlighted by [124] involves the response to food-related visual cues in the insula, which arises solely when the creature is hungry [126]. In this instance, the presumed neural representation of food cues and their valence is context-specific and inaccessible beyond this context; such as to depict food's absence or in situations where hunger isn't present.

Subsequently, the organism acquires the capacity to endogenously generate its pragmatic neural representations. This endogenous generation *detaches* and abstracts these representations from their initial sensorimotor context, divorcing them from mandatory links with external sensory inputs, action execution, or the organism's initial behavioral state that facilitated their emergence. However, these representations retain their original grounding: they are still *about* something, can be utilized to evaluate and select alternative actions (whether executed or imagined), and hold meaning and significance for the organism. Importantly, their detachment equips organisms with the cognitive capacity to envision and contemplate "what is not there"; be it potential actions, desired yet non-existent entities, or communicable concepts. Detachment thus fuels advanced cognitive functions, particularly future-oriented cognition like planning and imagination, potentially explaining the evolutionary benefits of detachment itself. Most cognitive theories suggest that genuine "mental life" arises when an organism can endogenously generate neural representations of the world; be it for shaping it under personal preferences or for contemplation. Expanding upon the example of food-related cues, detached representations enable pondering or conversing about food even in the absence of food or hunger. Additionally, knowledge gained about food (or navigation or similar subjects) could be abstracted to support novel capabilities, such as understanding narratives or envisioning rewarding journeys. These instances illustrate how detachment engenders a fundamental enhancement or shift from pragmatic to epistemic or descriptive objectives in neural representations. In essence, detached representation corresponds to the "descriptive" representations often associated with our mental experiences. And which become enriched through social interaction and linguistic communication [127–130].

Remarkably, this perspective unveils an intriguing bidirectional interplay between the two key functions of generative models that have been emphasized throughout this paper: engaging with the lived world and understanding it. Conventional cognitive theory often posits that, in order to interact effectively with the world, an organism must first achieve a level of understanding about it. Simultaneously, we've underscored here that comprehending the world – and generating meaning from it – necessitates active interaction. In active inference, these two processes are intimately entwined. In a manner analogous to how the minimization of variational free energy across short timescales sets in motion reciprocal loops between immediate perception and action, the same minimization across longer timescales fosters reciprocal loops between the act of understanding the world and actively engaging with it.

Numerous researchers have proposed diverse iterations of this narrative, yet much remains speculative. One challenge is reconciling the notion of detachment with a gradualist perspective of brain evolution. Future investigations should explore whether delineating the evolution of generative models for active inference across simple-to-complex organisms might elucidate the development of detached and future-oriented cognitive abilities from action-control loops [65].

## Acknowledgements

This research received funding from the European Union's Horizon 2020 Framework Programme for Research and Innovation under the Specific Grant Agreements No. 945539 (Human Brain Project SGA3) to GP, No. 952215 (TAILOR) to GP, and No. 951910 (MAIA) to IPS; the European Research Council under the Grant Agreement No. 820213 (ThinkAhead) to GP the Wellcome Centre for Human Neuroimaging (Ref: 205103/Z/16/Z) to KF, a Canada-UK Artificial Intelligence Initiative (Ref: ES/T01279X/1) to KF, and the MUR projects PE0000013-FAIR to GP, IR0000011–EBRAINS-Italy to GP and PRIN 2017KZNZLN to IPS. The funders had no role in study design, data collection and analysis, decision to publish, or preparation of the manuscript.

104. Kato S., H.S. Kaplan, T. Schrödel, *et al.* 2015. Global brain dynamics embed the motor command sequence of Caenorhabditis elegans. *Cell* 163: 656–669. https://doi.org/10.1016/j.cell.2015.09.034
105. Brennan C. & A. Proekt. 2019. A quantitative model of conserved macroscopic dynamics predicts future motor commands. *eLife* 8: e46814. https://doi.org/10.7554/eLife.46814
106. Attias H. 2003. Planning by Probabilistic Inference. In *Proceedings of the Ninth International Workshop on Artificial Intelligence and Statistics*.
107. Botvinick M. & M. Toussaint. 2012. Planning as inference. *Trends in Cognitive Sciences* 16: 485–488. https://doi.org/10.1016/j.tics.2012.08.006
108. Kaplan R. & K.J. Friston. 2018. Planning and navigation as active inference. *Biol Cybern* 112: 323–343. https://doi.org/10.1007/s00422-018-0753-2
109. Lanillos P., C. Meo, C. Pezzato, *et al.* 2021. Active Inference in Robotics and Artificial Agents: Survey and Challenges. . https://doi.org/10.48550/arXiv.2112.01871
110. Friston K.J. 2010. The free-energy principle: a unified brain theory? *Nat Rev Neurosci* 11: 127–138. https://doi.org/10.1038/nrn2787
111. Frank S.A. 2012. Natural selection. V. How to read the fundamental equations of evolutionary change in terms of information theory. *Journal of Evolutionary Biology* 25: 2377–2396. https://doi.org/10.1111/jeb.12010
112. Friston K., D.A. Friedman, A. Constant, *et al.* 2023. A Variational Synthesis of Evolutionary and Developmental Dynamics. *Entropy* 25: 964. https://doi.org/10.3390/e25070964
113. Vanchurin V., Y.I. Wolf, M.I. Katsnelson, *et al.* 2022. Toward a theory of evolution as multilevel learning. *Proceedings of the National Academy of Sciences* 119: e2120037119. https://doi.org/10.1073/pnas.2120037119
114. Albarracin M., D. Demekas, M.J.D. Ramstead, *et al.* 2022. Epistemic Communities under Active Inference. *Entropy* 24: 476. https://doi.org/10.3390/e24040476
115. Constant A., M.J. Ramstead, S.P. Veissiere, *et al.* 2018. A variational approach to niche construction. *Journal of The Royal Society Interface* 15: 20170685.
116. Laland K., B. Matthews & M.W. Feldman. 2016. An introduction to niche construction theory. *Evol Ecol* 30: 191–202. https://doi.org/10.1007/s10682-016-9821-z
117. Vasil J., P.B. Badcock, A. Constant, *et al.* 2020. A World Unto Itself: Human Communication as Active Inference. *Frontiers in Psychology* 11:.
118. Veissière S., A. Constant, M.J.D. Ramstead, *et al.* 2019. Thinking Through Other Minds: A Variational Approach to Cognition and Culture. *Behavioral and Brain Sciences*.
119. Corcoran A.W., G. Pezzulo & J. Hohwy. 2020. From allostatic agents to counterfactual cognisers: active inference, biological regulation, and the origins of cognition. *Biol Philos* 35: 32. https://doi.org/10.1007/s10539-020-09746-2
120. Bernardi S., M.K. Benna, M. Rigotti, *et al.* 2020. The Geometry of Abstraction in the Hippocampus and Prefrontal Cortex. *Cell* 183: 954-967.e21. https://doi.org/10.1016/j.cell.2020.09.031
121. Buzsáki G., A. Peyrache & J. Kubie. 2014. Emergence of Cognition from Action. *Cold Spring Harb. Symp. Quant. Biol.* 79: 41–50. https://doi.org/10.1101/sqb.2014.79.024679
122. Buzsáki G. & E.I. Moser. 2013. Memory, navigation and theta rhythm in the hippocampal-entorhinal system. *Nat Neurosci* 16: 130–138. https://doi.org/10.1038/nn.3304
123. Cisek P. 2019. Resynthesizing behavior through phylogenetic refinement. *Atten Percept Psychophys*. https://doi.org/10.3758/s13414-019-01760-1
124. Cisek P. 2021. An evolutionary perspective on embodiment. *Handbook of embodied psychology: Thinking, feeling, and acting* 547–571.
34